\documentclass[a4paper,12pt]{article}
\usepackage[dvips]{graphicx}
\begin{document}
\begin{center}
{\Large\bf Search for the $\Delta ^{++}$ component in $^{12}$C ground state
using $^{12}$C$(\gamma, \pi ^{+}p)$ reaction }\\

\vspace{3mm}
{\large A.I. Fix$^a$, I.V. Glavanakov$^b$, Yu.F. Krechetov$^b$\footnote{
Nuclear Physics Institute at Tomsk Polytechnic University, P.O. 
Box 25, Tomsk 634050, Russia. Tel. +7 3822 423994, Fax +4 3822 423934, 
E-mail krechet@npi.tpu.ru}
, \\O.K. Saigushkin$^b$,
E.N. Schuvalov$^b$, A.N. Tabachenko$^b$}\\

{\small\it $^a$ Tomsk Polytechnic University, 634050, Tomsk, Russia\\
$^b$ Nuclear Physics Institute
at Tomsk Polytechnic University,\\ 634050, Tomsk, Russia}
\end{center}

\begin{abstract}
The differential cross section for the reaction $^{12}$C$(\gamma ,\pi
^{+}p)$ has been measured in the $\Delta$ resonance region at
high recoil momenta of the residual nuclear system.
The data are analysed under the assumption that the
formation of $\pi^+p$ pairs may be interpreted as
$\gamma\Delta^{++}\rightarrow\pi^+p$ process,  proceeding on the
$\Delta^{++}$ constituent in a target nucleus.
As a result, the estimation $N_\Delta=0.028\pm 0.008$ deltas per nucleon 
in $^{12}$C was obtained.
\vspace*{3mm}

\end{abstract}

The interaction of energetic projectile particles with nuclei  
at high momentum transfer implies probing the nuclear
structure at short distances. A wide variety of developments
in the short range nucleon-nucleon dynamics concerns the internal
nucleon degrees of freedom. The nontrivial substructure of nucleon manifests
itself in existence of the internally excited states of nucleon 
i.e.\ baryon resonances or isobar. Therefore, in  modern nuclear models 
the conventional nuclear wave function consisting of nucleons only is
supplemented by the exotic components, the so-called isobar configurations. 
The role of nuclear isobar configurations in nuclear reactions involving high 
momentum transfer is one of the most interesting questions of medium energy
physics (see e.g.\ reviews \cite{Green76, Aren78}).

It was found within the frame of different approaches that the
strongest isobar admixture in nuclei stems from the
$\Delta$(1232)-resonance. Most theoretical and experimental investigations
of the $\Delta$-isobar configuration have been done for the deuteron,
$^3$He and $^3$H and to a smaller degree for havier nuclei. At the same
time, the amount of virtual deltas for more massive nuclei with their 
higher density is expected to be more essential than in lighter
nuclei. Here we would like to mention the 
explicit calculations of the $\Delta$ probability 
in $^4$He and $^{12}$C \cite{Horl78} that give estimates of about
4.5$\%$ and 3.2$\%$, respectivelly. The examinations of the 
existence of virtual $\Delta$s  
had little success despite significant experimental efforts. 
As for the experimental search for the isobar
admixture in nuclei with $A>4$, the observation 
of $\Delta^{++}$ knock-out from
$^9$Be in p$^9$Be collisions \cite{ABC94}
and the recent double charge
exchange experiment A$(\pi^+,\pi^-)$  on $^{12,13}$C, $^{90}$Zr, 
$^{208}$Pb \cite{Morr98} present the most interest.
The corresponding data analysis
point to the probable existence of the virtual $\Delta$s in p-shell nuclei 
at the level of $1\div3\%$.
The photo- and electroproduction processes as a test of 
$\Delta$ admixture in nuclei were considered in Ref.\ \cite{Lipkin}.
Laget proposed the combined study of $(e,e'\Delta^{++})$ and 
$(e,e'\Delta^o)$ reactions on $^{3}$He \cite{Laget}.
The coresponding investigations are performed now 
at MAMI and Jefferson Lab.\ \cite{Kuss, Berman}.
It is noted in Ref.\ \cite{blomq}, that results for the reaction 
$^3$He$(e,e'\pi^{\pm})$ obtained at MAMI do not contradict 
with the assumption of a preformed $\Delta^{++}$.
As for the reactions with real photons, only one 
measurement of $^{12}$C$(\gamma,\pi^+ p)$ reaction has been reported 
in Ref.\ \cite{McKenz97}. The estimation $\Delta$ admixture 
in $^{12}$C from these data is difficult since the experimental cross
section was averaged over the wide range of the
angles of emitted particles.

In this paper we present the experimental results for
the differential cross section of the reaction
\begin{equation}\label{gammaA}
\gamma + ^{12}C \rightarrow  \pi^+p + X\,.
\end{equation}
The measurements are performed in the kinematical region of high
momenta transferred to the residual nuclear system $X$. Our investigations
are designed to determine the admixture of $\Delta(1232)$-component in
the $^{12}$C ground state. We analyse the data under the assumption
that $\pi^+p$ pairs are the product of the
$\gamma\Delta^{++}\rightarrow\pi^+p$ process which proceeds on the
$\Delta^{++}$-constituent preexisting in a target nucleus.
The method of using reactions of the type (\ref{gammaA}) also examined
in our previous paper \cite{FGK99} has two main advantages.
First, since the
$\Delta^{++}$ cannot be excited by photons on a single nucleon the
interpretation of measurements is free from the difficulty to
distinguish the knocked-out $\Delta$s from those created in the
reaction. Second, since virtual $\Delta$s are produced in the NN
collisions, it is reasonable to expect that the probability for finding
$\Delta$ in a nucleus is quadratic in nuclear density. Therefore,
the search for the $\Delta$-components in electromagnetic reactions is
preferred since the photon beam is able to probe the whole nuclear
wolume up to the region of highest densities.  As a consequence, more
direct test of isobar configuration may be carried out.
This distinguishes reactions (\ref{gammaA}) from those using
strongly interacting probes, which undergo multiple scattering
before reaching central-nuclear densities. 

Measurements of the differential cross section of the reaction
$^{12}$C$(\gamma ,\pi ^{+}p)$ were performed at the Tomsk synchrotron.
The bremsstrahlung photon beam have been produced by electrons with
energy 500 and 420 MeV. The experimental
setup is shown in Fig.\ 1. It includes
two channels for detecting the charged pion and
the proton in coincidence.

The pions with 181 Mev/c mean momentum were detected by a strongly
focusing magnetic spectrometer at the angle $54^{o}$ with respect to the
photon beam.  The angular and momentum acceptances were $3\cdot
10^{-3}$ sr and $24\%$, respectively. It was determined by the telescope
that includes two scintillation counters. 
The momentum resolution of $1\%$ was
available by use of the scintillation hodoscope
\cite{PR85IFVE} located in the focal plane of the spectrometer.
\begin{figure}[h]
\unitlength=1cm
\centering
\begin{picture}(10,5)
\includegraphics[width=10cm,keepaspectratio]{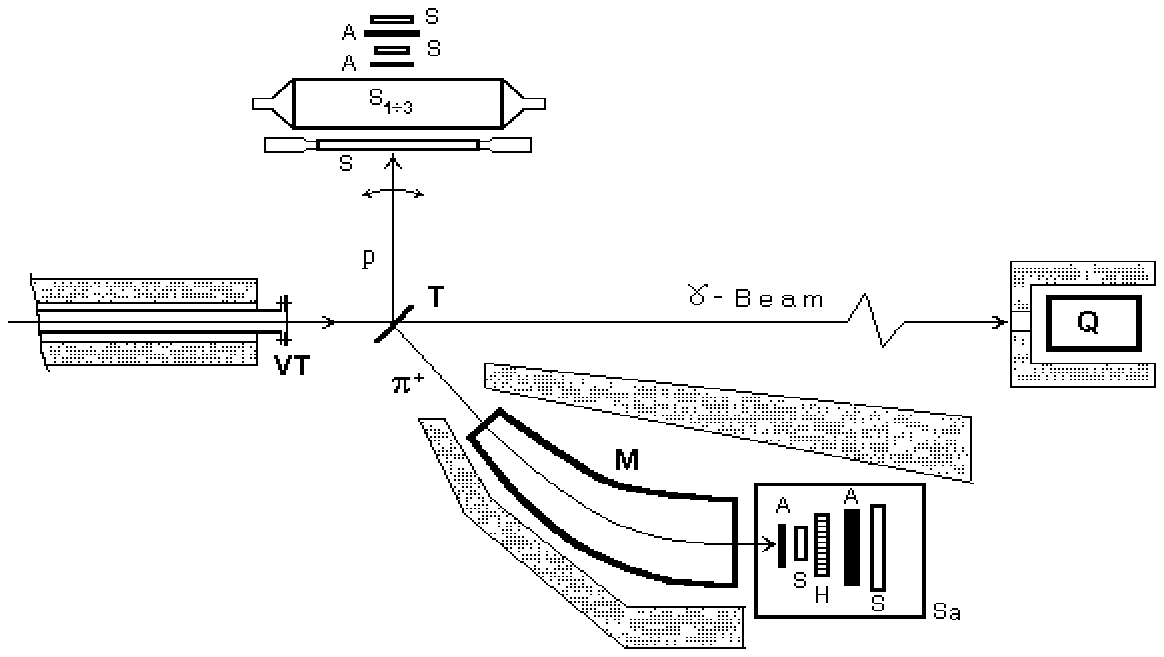}
\end{picture}
\vspace*{7mm}
\caption{ Layout of the experimental setup: (T) target, (VT) vacuum
box, (Q) Gauss-quantameter, (AM) analyzing magnet, (S) scintillation
counters,  (H) hodoscope, (A) absorbers.}
\end{figure}
Proton channel includes ($\Delta E,  E $) - system
of the scintillation counters on the polystyrene base
and two auxiliary scintillation counters with absorbers intended
for the  energy calibration and for monitoring the
stability of the ($\Delta E,  E $) - system.
The solid angle of the proton channel
defined by the $\Delta E$-counter was equal to
0.26 sr. The mean polar angle $\theta_p$  with respect to the
photon beam  and the angular coverage of the proton channel
were equal to $75^o$ and $\pm 19^o$, respectively.
The $E$-detector includes three scintillation counters of dimensions
$10\times 10\times 50$ cm$^3$, which were disposed on the top of
one another. In the region of the proton energy 
$T_p = 40\div 120$ MeV the analysis of the impulse amplitudes from
the $\Delta E$- and $E$-counters allows us to determine the polar angle
and the energy of the registered proton with accuracies better 
than $\sigma_\theta = 3^o$ and $\sigma_E = 4$ MeV, respectively. 
The accuracy of the azimuthal angle measurement
determined by the counter dimensions was $ \sigma_\phi \sim
2^o$. The auxiliary scintillation counters chose narrow-direct 
beam of the particles. 
The pions and protons of the beam with the minimal energy
were used for the energy calibration of the proton channel.
This minimal energy was determined by the material thickness 
of the proton channel. 
We carried out the calibration 
and the major run simultaneously. 

To reduce the cosmic background, the pion channel detectors
were covered with the big-area scintillation counter Sa functioning in
the anticoincidence condition. With the similary purpose the final trigger
was formed only during the accelerator radiation impulse.

The target was a carbon slab of the natural isotope composition
$4.35\cdot 10^{22}$ nuclei/cm$^{2}$ in thickness. The total energy of
the photon beam was measured by the Gauss-quantameter $\cite{quant}$.

In the present experiment 45 events of the $\pi^+
p$-coincidences have been yielded. Figure 2 shows the distribution of
events as a function of the proton emission angle and the proton energy
for two accelerator electron energy.

\begin{figure}
\unitlength=1cm
\centering
\begin{picture}(6,6.4)
\includegraphics[width=6cm,keepaspectratio]{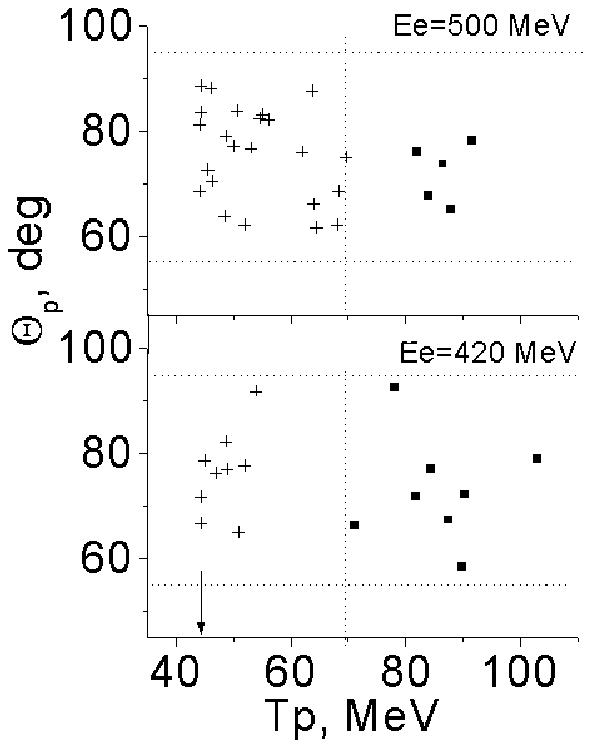}
\end{picture}
\caption{Experimental events distribution versus proton emission angle and
proton kinetic energy for two accelerator electron energy. Crosses are the
background events. Solid squares are the events selected for the
differential cross section estimation. Arrow is the detection threshold.
The horizontal dashed curves restrict the angle range of the
detection protons. The verticalal dashed curve is the cut of the
proton energy range.}
\end{figure}

The events are concentrated in the
vicinity of two proton energies 50 MeV and
85 MeV. This effect can readily be seen
on the lower panel corresponding to the electron energy
$E_e$ = 420 MeV. At this energy two
$\pi^+ p$ production mechanisms may be recognized. The first
mechanism is caused by the binary process
$\gamma\Delta^{++}\rightarrow\pi^+p$, being the subject of the present
investigation.
The second one results from the $\pi^o p$- and
$\pi^+ n$-pairs production with following charge-exchange
rescattering of neutral pions and neutrons. It is reasonably to expect,
that the events stemming from the $\pi^o p$- and $\pi^+ n$-pairs
production are mainly grouped in the region of small proton energies.
This is due to the fact that the processes of quasifree $\pi^{+}n$ and
$\pi^o$p production dominate in the region of small momenta of the residual
nuclear system. In the kinematical conditions considered in our 
experiment, this region corresponds to the small proton energies.
Therefore, we assume that for the energies $T_p\le$ 55 MeV 
both mechanisms mentioned above are significant.
At the same time, in the region $T_p\ge$\ 70 MeV where the intencity of
the quasifree pion photoproduction decreases in approximatelly
$10^2$ times, the background effects are of minor importance and
the registered events must be due mainly to the direct $\pi^+p$
formation on preexisting deltas.

At the energy $E_e$ = 500 MeV, one can see the higher
density of events corresponding to the proton energies less than 70
MeV. Here the events due to the
$\pi^+p$-pairs formation in the reaction
$^{12}$C$(\gamma,\pi^+\pi^-p)^{11}$B are expected to give some
contribution. The estimation of this background has been
done in the framework of the quasifree approximation.
The corresponding formalism is presented in detail
in Ref. \cite{pipi}.
The calculation showes that in the region $T_p\approx 55\div 70$ MeV
the experimental yield is due practically to the
two-pion production mechanism, while for the proton energies $T_p\ge$ 75 MeV
the net effect from
this process is estimated to be less than $1\%$.
The above mentioned grounds led us to select for the analysis only the
events, which lay within the interval $T_p=70 \div 110$ MeV.

We write the differential cross section as

$$
 \frac{d^3\sigma}{ dE_pd\Omega_pd\Omega_\pi}=
\frac{N_{\pi p}E_{\gamma max}}{\Delta E_\pi \, \Delta \Omega_\pi \,
                    \Delta E_p \, \Delta \Omega_p \, W_\gamma \, t}
\biggm/f(E_\gamma) \, \Biggl|
  \frac{\partial E_\gamma}{\partial E_\pi} \Biggr |,
$$
\noindent
where $E_\gamma, E_\pi$, and $E_p $  are the energies of the
photon, pion, and proton, respectively; $N_{\pi p}$ is the
number of events in the phase space defined by the intervals
$ \Delta E_\pi$, $\Delta
E_p$, $ \Delta \Omega_\pi$, $\Delta \Omega_p $; $W_\gamma$  is the  total
energy of the photon beam; $t$ is the target thickness;
$ E_{\gamma max} $ is the endpoint energy of the bremsstrahlung spectrum
$ f(E_\gamma) $, which is normalized as
$$
 \int_0^{E_{\gamma max}} f(E_\gamma) \,E_\gamma \ dE_\gamma =
   E_{\gamma max}.
$$
The kinematical quantities not measured in the reaction were determined
by solving the set of kinematical equations under the assumption that
the residual nucleus $^{11}$Be is in its ground state.

The cross section was averaged in the intervals
$E_\pi = 211\div 246$ MeV, $\theta_p = 56\div 94^o$,
$E_p = 70\div 110$ MeV. The mean photon energy for the events laying 
in the kinematical region under consideration was 345 MeV.
Within the procedure described above we obtain the following result

\begin{equation}\label{dsigma}
\frac{d^3\sigma}{ dE_pd\Omega_pd\Omega_\pi} = 8.5 \pm 2.5 
\frac{nb}{MeV sr^2}\,\, ,
\end{equation}
where the measurement error is statistical.

Now we recall briefly the method that was used in our calculation.
As has been noted, the elementary 
process $\gamma\Delta^{++}\rightarrow
\pi^+p$ was assumed to be the only reaction mechanism.  
We employ the impulse approximation and use the closure relation 
when summing over the 
states of the residual undetected nuclear system. This
gives  the differential cross section of the reaction 
$^{12}$C$(\gamma,\pi^+p)$ in lab system as
$$
\frac{d^3\sigma }{dE_p d\Omega _p d\Omega _\pi}=\frac{E_f \, E_p \,p_p\,
p_\pi^3\,  }
{4(2\pi )^5E_\gamma \mid E_f p_\pi^2 - E_\pi {\bf p}_\pi \cdot {\bf p}_f \mid }
\, f_\pi\, f_p \,\,\rho _{\Delta ^{++}}({\bf
p}_\Delta )\,
\frac12\sum_{\lambda =\pm 1} \overline{\mid {t_\lambda }\mid ^2}.
$$
Here letters $\gamma$, $\pi$, p, $\Delta$ and $f$ stand for the photon, pion,
proton, $\Delta$ and final nucleus respectively. The total energies and
momenta of the participating particles are denoted by $E_{i}$
and ${\bf p}_{i}$, $f_{\pi}$ and $f_{p}$
are the attenuation factors taking into account the 
absorption of the produced
pions and protons in nuclear medium (see e.g.\ \cite{lage})
;$\lambda$ is a index of the photon polarization; $t_{\lambda}$ is the 
elementary $\gamma \Delta^{++}\rightarrow \pi^{+}p$ 
amplitude; the function $\rho _{\Delta^{++}}
({\bf p}_\Delta)$ has the meaning of delta momentum distribution
in the ground state of initial nucleus.

The elementary $\gamma \Delta^{++}\rightarrow \pi^{+}p$ amplitude
$t_{\lambda}$ was obtained within the diagramatic approach. 
The corresponding
formulas are presented
in our previous work \cite{FGK99}.
Different terms of the amplitude are shown in Fig. 3\,(a - e).

\begin{figure}[t]
\unitlength 0.22mm
\begin{picture}(550,600)(0,0)

\multiput(45,602)(8,-8){10}{\oval(8,8)[lb]}
\multiput(45,594)(8,-8){10}{\oval(8,8)[rt]}
\put(45,514){\line(1,0){76}}
\put(45,508){\line(1,0){76}}
\multiput(151,522)(21,21){4}{\line(1,1){18}}
\put(151,508){\line(1,0){76}}
\put(136,514){\circle{30}}
\put(0,615){$\gamma\,(E_\gamma,{\bf p}_\gamma\,)$}
\put(185,615){$\pi^+\,(E_\pi,{\bf p}_{\pi}\,)$}
\put(0,480){$\Delta^{++}(E_\Delta,{\bf p}_\Delta)$}
\put(185,480){$p\,(E_p,{\bf p}_p\,)$}
\put(265,540){\Large =}

\multiput(360,602)(8,-8){7}{\oval(8,8)[lb]}
\multiput(352,602)(8,-8){8}{\oval(8,8)[rt]}
\put(360,544){\line(1,0){112}}
\put(360,540){\line(1,0){170}}
\multiput(475,546)(21,21){3}{\line(1,1){18}}
\put(420,542){\circle*{8}}
\put(471,542){\circle*{8}}
\put(429,500){(a)}
\put(550,540){\Large +}

\multiput(80,424)(8,-8){10}{\oval(8,8)[lb]}
\multiput(80,416)(8,-8){10}{\oval(8,8)[rt]}
\multiput(98,346)(21,21){4}{\line(1,1){18}}
\put(45,344){\line(1,0){112}}
\put(45,340){\line(1,0){170}}
\put(100,343){\circle*{8}}
\put(156,341){\circle*{8}}
\put(114,300){(b)}
\put(270,340){\Large +}

\multiput(410,424)(8,-8){10}{\oval(8,8)[lb]}
\multiput(410,416)(8,-8){10}{\oval(8,8)[rt]}
\multiput(432,346)(21,21){4}{\line(1,1){18}}
\put(360,344){\line(1,0){70}}
\put(360,340){\line(1,0){170}}
\put(428,343){\circle*{8}}
\put(490,341){\circle*{8}}
\put(434,300){(c)}
\put(550,340){\Large +}

\multiput(78,202)(8,-8){7}{\oval(8,8)[lb]}
\multiput(70,202)(8,-8){8}{\oval(8,8)[rt]}
\multiput(130,146)(21,21){3}{\line(1,1){18}}
\put(45,108){\line(1,0){85}}
\put(45,104){\line(1,0){170}}
\multiput(130,110)(0,10){4}{\line(0,1){8}}
\put(130,107){\circle*{8}}
\put(130,146){\circle*{8}}
\put(270,140){\Large +}
\put(118,65){(d)}

\multiput(368,192)(8,-8){10}{\oval(8,8)[lb]}
\multiput(368,184)(8,-8){10}{\oval(8,8)[rt]}
\multiput(445,110)(21,21){4}{\line(1,1){18}}
\put(360,110){\line(1,0){85}}
\put(360,106){\line(1,0){170}}
\put(446,108){\circle*{8}}
\put(433,65){(e)}
\end{picture}
\vspace{-1cm}
\caption{Diagrams for the $\gamma \Delta^{++}\rightarrow \pi^{+}p$  amplitude
used in the present calculation: (a) s- chanal term,  (b), (c) u-chanal terms,
(d) pion pole term, (e) seagull term.}
\end{figure}
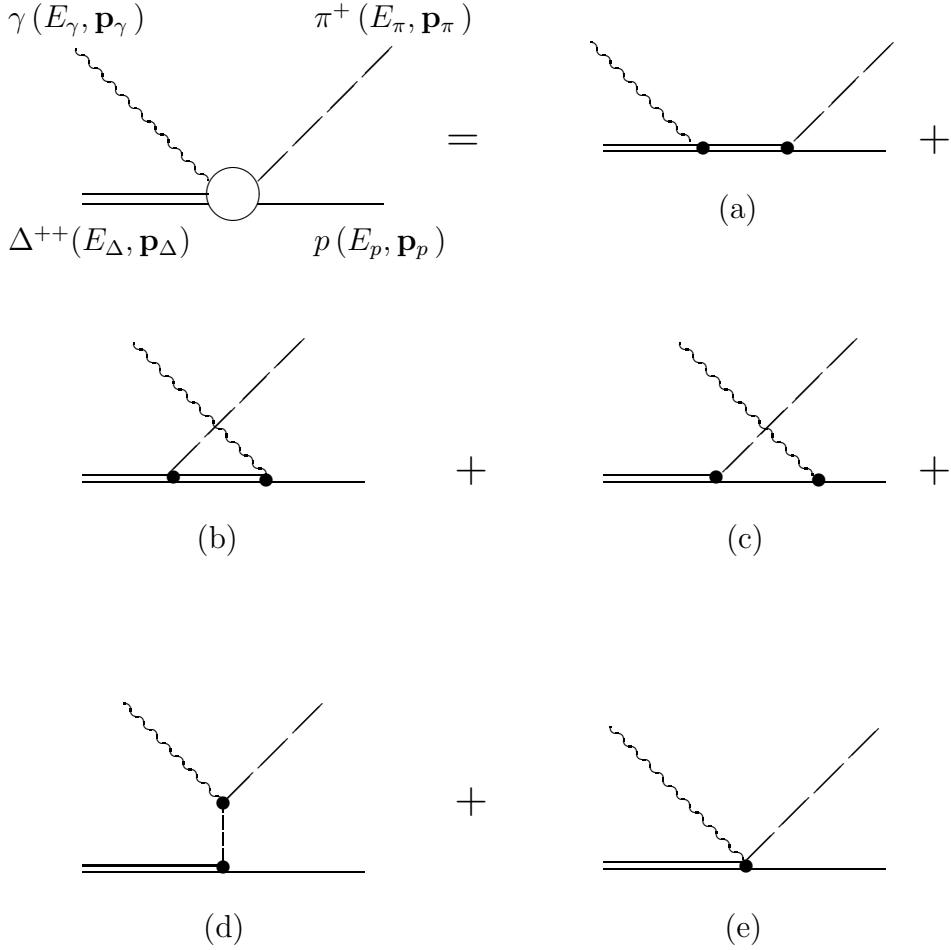
We include in the matrix element the new term (3.b) which was missed in the 
previous analysis \cite{FGK99}. 
As the calculation shows, this term is of minor importance
when embedding the 
$\gamma \Delta^{++}\rightarrow \pi^{+}p$ transition operator
into the nuclear process and does not visibly influence our prevoius results.
In the nonrelativistic limit up to the order 
$(p/M)^2$ the elementary amplitude may be written as
$$
t_\lambda = i\,
\frac{\sqrt{2}\,f_{\pi N\Delta }}{m_\pi }\frac e{2\,M_p}
\left\{
2\stackrel{+}{\bf S}
\cdot\,{\bf p}_{\pi p}
\frac{2\,{\bf p}_\Delta \cdot\,{\mbox{\boldmath$\varepsilon$}}_{\lambda}
\, + i\,\displaystyle \frac{\mu_{\Delta^{++}}}{3} 
\,{\mbox{\boldmath$\sigma$}}_{\Delta} \cdot\,[{\bf p_\gamma}\times 
{\mbox{\boldmath$\varepsilon$}}_{\lambda} ]}
{E_\Delta +E_\gamma-E_\Delta^{\prime}+i\,\Gamma_\Delta / 2}  \right.
$$
$$
+\,i\,\frac{f_{\pi \Delta \Delta }}{f_{\pi N\Delta }}\,\frac{3\,\mu _{N\Delta }
\stackrel{+}{\bf S}
\cdot\, [{\bf p}_{\gamma \Delta }\times {{\mbox{\boldmath$\varepsilon$}
}_{\lambda}]}}
{E_\Delta -E_\pi
-E_{\Delta}^{\prime \prime }}\,{\mbox{\boldmath$\sigma$} }_\Delta\cdot{\bf
p}_\pi + 2\,\frac{{\bf p}_p  \cdot\,{\mbox
{\boldmath$\varepsilon$}}_{\lambda}\, +i\,\displaystyle{\frac{\mu
_p}2}\,{\mbox{\boldmath$\sigma$}}\cdot\,[{\bf p_\gamma\times
{\mbox{\boldmath$\varepsilon$}}_{\lambda} ]}}
{E_\Delta-E_\pi -E_p^{\prime }}\stackrel{+}{\bf S}\cdot\,{\bf p}_{\pi p} 
$$
$$
\left.
-\, \frac{4\,M_p}{t-m_{\pi}^2}\stackrel{+}
{\bf S}\cdot\,({\bf p}_\pi-{\bf p}_\gamma)
\,{\bf p}_\pi\cdot\,{\mbox{\boldmath$\varepsilon$}
}_{\lambda}\, 
-\,2\,M_p \stackrel{+}{\bf S}\cdot\, 
{\mbox{\boldmath$\varepsilon$}}_{\lambda} \right\}
$$
For the hadronic coupling constants 
we employ $f_{\pi N \Delta}^2/4\pi$=0.37 from decay 
$\Delta \rightarrow \pi N$ and $f_{\pi \Delta 
\Delta}=4/5 f_{\pi NN}$, as predicted by the trivial quark model.
The magnetic moments used in the calculation are
$\mu _{p}$\,= 2.79, $\mu_{\Delta^{++}}$\,= 4.3 and $\mu_{N\Delta}$\,= 3.24
in terms of nuclear magnetons. The variables $E'_p=M_p+
({\bf p}_\Delta-{\bf p}_\pi)^2/2M_p$, $E''_{\Delta}=M_{\Delta}+ 
({\bf p}_\Delta-{\bf p}_\pi)^2/2M_{\Delta}$ and 
$E'_{\Delta}=M_{\Delta}+
({\bf p}_\Delta+{\bf p}_\gamma)^2/2M_{\Delta}$
are the energies of the intermediate nucleon and deltas;
${\bf p}_{\pi p}=({\bf p}_\pi M_p-{\bf p}_pE_{\pi})/(M_p+E_{\pi})$
and
${\bf p}_{\gamma \Delta}=({\bf p}_\gamma M_{\Delta}-({\bf p}_{\Delta}-{\bf p}_\pi)
E_{\gamma})/(M_{\Delta}+E_{\gamma})$ are the
relative momenta in the $\pi p$ and $\gamma \Delta $ systems;
$\bf {S}$ is transition operator between the states with spin 3/2 and 1/2; 
$\mbox{\boldmath$\sigma$}_\Delta$
is the analog of the Pauli spin matrix for the spin $\frac32$ object
(concrete representations 
of $\bf S$ and $\mbox{\boldmath$\sigma$}_\Delta$
are given e.g.\ in \cite{Aren78}).

The delta momentum distribution 
$\rho _{\Delta ^{++}}({\bf p})$ obeys the following normalization  
condition
\begin{equation}\label{norm}
\int{\rho_{\Delta^{++}}({\bf p})\frac{
d{\bf p}}{(2\pi)^3}}=\frac{A N_\Delta^c}{4}\,\, ,
\end{equation} 
where $N_\Delta^c$ is the number of deltas per nucleon
in the ground state of the nucleus $^{12}$C and A = 12 is the mass number. 
This function was calculated as 
\begin{equation}\label{rho}
\rho_{\Delta^{++}}({\bf p})=4\cdot\frac43 \pi 
R^{3}\ n_\Delta^c({\bf p}), 
\end{equation} 
where $R=3.2$ fm is the square-well radius and
$n_\Delta^c({\bf p})$ is the $\Delta$'s occupation number
in $^{12}$C. For lack of available analysis for the function
$n_\Delta^c({\bf p})$ (or $\rho_{\Delta^{++}}({\bf p})$)
for p-shell nuclei, we
use the results of Ref.\ \cite{Cenni89} where the $\Delta$'s
occupation number inside the nuclear matter 
$n_\Delta^m({\bf p})$ is presented.  In order to relate 
$n_\Delta^c({\bf p})$ to $n_\Delta^m({\bf p})$, we
assume that the momentum distribution $\rho_{\Delta^{++}}({\bf
p})$ of deltas in $^{12}$C is proportional to that for the
nuclear matter. This leads to 
\begin{equation}
n_\Delta^c({\bf p})=\frac{N_\Delta^c}{N_\Delta^m}\:\left(
\frac{r_0^m}{r_0^c} \right)^3\:
n_\Delta^m({\bf p}). 
\end{equation}
In the actual calculation  
we use the following radial parameters ($r_0=RA^{-1/3}$) \cite{Bethe}
\begin{equation}
r_0^m=1.12\ \mbox{fm}, \quad r_0^c=1.4\ \mbox{fm}.
\end{equation} 
Treating $N^c_\Delta$ as a free parameter we
determine its value by fitting our calculation
to the experimental result (\ref{dsigma}).
Taking $N_\Delta^m$=0.07 from \cite{Cenni89} this 
approach gives 
\begin{equation}
N_\Delta^c=0.028\pm 0.008  \ \mbox{deltas per nucleon}\, ,
\end{equation}
where indicated uncertainty is statistical only.
This value is in rough agreement with the results obtained by other
authors for p-shell nuclei \cite{ABC94,Morr98}.
 
The differenial cross section as function the
energy of the proton $E_{p}$ receivied from data for $E_{e}$= 500 MeV and
$E_{e}$=420 MeV as mean values in the intervals $\Delta \theta_{p}=56^{0}\div 94^{0}$,
$\Delta E_{\pi}=211\div 246$ MeV are given in Fig.4
The differential cross  section calculated in our model with
$N_{\Delta}^{^{12}C}$=0.028 are also shown.

\begin{figure}[h]
\unitlength=1cm
\centering
\begin{picture}(7,7)
\includegraphics[width=7cm,keepaspectratio]{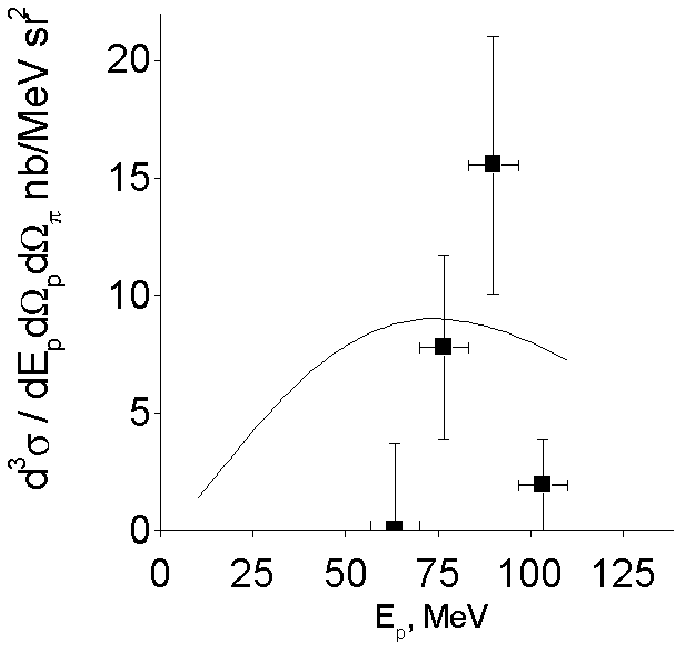}
\end{picture}
\caption{The differential cross section of the the $^{12}C(\gamma ,\pi ^{+}p)$
reaction at mean $E_{\gamma}$=340 MeV as a function of the kinetic energy
$T_{p}$ for the intervals $\Theta_{p}=56^{0}\div 94^{0}$,
$E_{\pi}=211\div 246$ MeV. The solid curve is the risult of the our the
calculations.}
\end{figure}

It must be kept in mind that our quantitative conclusion depends
strongly on the model adopted for the $\Delta$ dynamics in
nuclei. This question is also briefly considered in \cite{FGK99}.
Thus it is highly desirable to have more realistic model for the
delta momentum distribution in $^{12}$C. 

This work was supported by the Russian Foundation for Basic Research
under the Contracts No. 96-02-16742, No. 97-02-17765, and No. 99-02-16964.

\end{document}